\documentclass[aps,pre,reprint,amsmath,amssymb]{revtex4-2}
\usepackage{graphicx,subfigure}
\usepackage{dcolumn}
\usepackage{epstopdf}
\usepackage{color}
\usepackage{upgreek}
\usepackage[hidelinks]{hyperref}

\usepackage{bm}

\makeatletter
\def\btt#1{\texttt{\@backslashchar#1}}%
\DeclareRobustCommand\bblash{\btt{\@backslashchar}}%
\makeatother

\begin{document}

\title{ Interactions of charged microrods in chiral nematic liquid crystals}

\date{\today}
\author{Muhammed Rasi M$^{1}$, Ravi Kumar Pujala$^{2}$, Sathyanarayana Paladugu $^{2}$,  and Surajit Dhara$^{1}$}
\email{surajit@uohyd.ac.in} 
\affiliation{$^1$School of Physics, University of Hyderabad, Hyderabad-500046, India\\
$^{2}$Department of Physics, Indian Institute of Science Education and Research, Tirupati, Andhra Pradesh 517507, India
}

\begin{abstract}
 We study the pair interaction of charged silica microrods in chiral nematic liquid crystals and show that the microrods with homeotropic surface anchoring form a bound state due to the competing effect of electrostatic (Coulomb) and elastic interactions. The robustness of the bound state is demonstrated by applying external electrical and mechanical forces that perturbs their equilibrium position as well as orientation. In the bound state we have measured the correlated thermal fluctuations of the position, using two-particle cross-correlation spectroscopy that uncovers their hydrodynamic interaction. These findings reveal unexplored aspects of liquid-crystal dispersions which are important for understanding the assembly and dynamics of nano and microparticles in chiral nematic liquid crystals.
\end{abstract}

\preprint{HEP/123-qed}
\maketitle

\section{Introduction}
   Microparticles in nematic liquid crystals (NLCs) create spatial deformation of the director field $\bf{\hat{n}}$ (mean molecular orientation), and induce topological point or line defects~\cite{Igor, Stark,Ivan}. Spherical microparticles with homeotropic anchoring mainly stabilize dipolar and quadrupolar type director distortions such that the particles are accompanied by either a hyperbolic hedgehog or a disclination ring (Saturn ring). The defects and the ensuing elastic distortion mediated long-range interactions have created immense interests in recent years owing to the potential application for designing two and three dimensional colloidal crystals~\cite{Igor1, Musevic}. 
When the dispersing medium is a twisted nematic or chiral nematic, the director twists in space, and the spherical microparticles exhibit a number of interesting defect structures wherein a single defect loop wraps around the microsphere, with the winding layout being more complex in cells with a higher twist. The interaction of such defect loop decorated  microspheres often leads to metastable  {\cite{Jampani} and reconfigurable states due to the screening of elastic interaction by the periodic structure of the chiral nematics ~\cite{Jam1, Tkalec1, Miha}.} Other types of defect in chiral nematic includes vortex-like nonsingular defects~\cite{Poulin,Tkalec}, knots and links~\cite{Tkalec1}, torons~\cite{Evans},  skyrmions~\cite{Ivan1,Pos} etc. 

 The nontrivial shapes of the particles also greatly influence the elastic distortion and the ensuing interaction. For example, anisotropic particles such as the microrods~\cite{rod1,16,svb,dam,mni}, nanorods~\cite{Rasi,frh}, star~\cite{17}, bullet~\cite{18}, ellipsoidal~\cite{mta},  fractal~\cite{19} and peanut-shaped~\cite{ours} particles have been reported to exhibit complex defect structure and assembly in nematic LCs. In chiral nematics the defects and elastic interaction of anisotropic particles are obviously going to be  more complex and such studies are largely unexplored~\cite{21}.

 Apart from the elastic interaction, electrostatic (for charged particles) and hydrodynamic interactions of microparticles in liquid crystals play an important role in their assembly. Recently some interesting aspects of electrostatic interaction of particles in nematic LCs have been experimentally demonstrated~\cite{ivan1,ivan2,ivan3}. Theoretically hydrodynamic effects in LC colloids have beenstudied~\cite{lintvuri1,lintvuri2,lintvuri3} but not directly explored experimentally.
 The hydrodynamic interaction has been studied in aqueous colloidal systems by optically trapping two Brownian particles using laser tweezers~\cite{Jens,Bar}. However, the particles in LCs  can not be directly trapped by laser tweezers in the conventional ways~\cite{Skarabot, Musevic2}, thus making the direct measurement of hydrodynamic interaction problematic. 

\begin{figure}[!ht]
\center\includegraphics[scale=0.22]{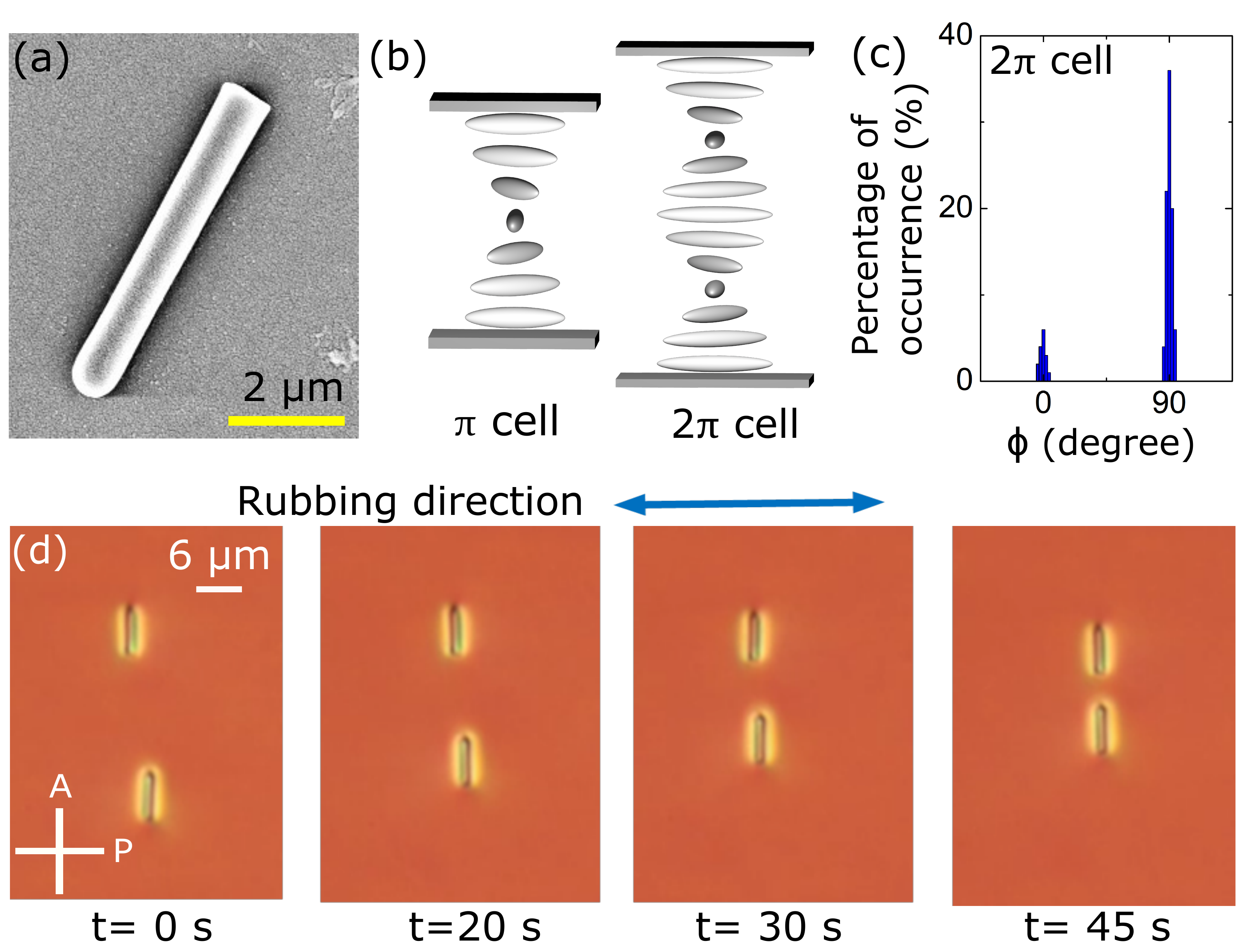}
\caption{(Color Online) (a) Field emission scanning electron microscope (FESEM) image of a silica microrod. (b) Director twist in $\pi$ and 2$\pi$-twisted nematic cells. (c) Percentage of microrods oriented parallel ($\phi=0^{\circ}$) and perpendicular ($\phi=90^{\circ}$) to the rubbing direction in a 2$\pi$-twisted cell. (d) Polarising optical microscope images of two interacting microrods with elapsed time (Movie S1)\cite{sup}. Horizontal arrow indicates the rubbing direction. P, A indicates polariser and analyser. 
 \label{fig:figure1} }
\end{figure}
In this paper we report experimental studies on charged silica microrods with homeotropic anchoring condition in chiral NLCs and show that for a particular orientation a pair of particles are self-trapped due to the competing effect of electrostatic  and elastic pair interaction, leading to a bound state, in which they are separated by a certain distance. 
 We show that the trapping is robust against external perturbations that tend to change their position as well as orientation. In the bound state two microrods interact through hydrodynamic forces. 

\section{Experiment}

We synthesized silica microrods by using a wet chemical method following the procedure described by Kujik \textit{et al.}~\cite{Kuijk,Rasi}. To begin with, 3 g of PVP (Polyvinylpyrrolidone) was dissolved in 30 ml of 1-pentanol. After complete dissolution of PVP, 3 ml of ethanol (100\%), 0.84 ml Milli Q water and 0.2 ml aqueous sodium citrate dihydrate (0.17 M) were added and mixed thoroughly. Then 0.3 ml of TEOS (98\%) was added to the mixture and the reaction was left to continue for 24 hours. The product mixture was then centrifuged and fractionated to obtain desired aspect ratio of microrods.
 The average length and diameter are 6~$\upmu$m and  800 nm, respectively, and one end of the microrods is round-shaped as shown in  Fig.\ref{fig:figure1}(a).
The microrods were coated with octadecyldimethyl-3-trimethoxysilyl propyl-ammonium chloride (DMOAP) before dispersing into the chiral NLCs for obtaining perpendicular or homeotropic anchoring of the director $\bf{\hat{n}}$. 
The chiral NLCs were prepared by doping a chiral agent; 4-cyano-40-(2-methylbutyl)-biphenylene (CB15) in 4-n-pentyl-4-cyanobiphenyl (5CB) nematic liquid crystal. The concentration of CB15 was adjusted to vary the helical pitch, which was measured using a wedge cell {\cite{Podolskyy}}. \\

The microrod-LC dispersions were studied in planar cells, which are prepared by two parallel glass plates (15 mm$\times$10 mm) coated with the polyimide AL-1254 (Nissan Chemicals). The plates were rubbed in an antiparallel way before assembling so that the helix axis of the chiral NLC is perpendicular to the plane of the substrate as shown in Fig.\ref{fig:figure1}(b). The pitch was adjusted in a cell of gap $d$ such that the nematic structure is twisted by multiples of half-pitch i.e., $d=N(p/2)$, which is denoted here as $N\pi$-twisted cell (Fig.\ref{fig:figure1}(b)). 
We used  $d=4~\upmu$m and $p=8~\upmu$m for $\pi$-twisted and  $d=5~\upmu$m and $p=5~\upmu$m for $2\pi$-twisted cells.
An infrared laser tweezers setup (Aresis, Tweez 250Si) attached to an inverted optical polarising microscope (Ti-U, Nikon)  was used in the experiments~\cite{Zuhail,Zuhail1,Rasi}. The laser trap movement was controlled by an acousto-optic deflector (AOD) and computer. The motion of the microrods was video recoded using a 60X water immersion objective (NIR Apo, Nikon) and a charge-coupled device (CCD) camera (iDs-UI) at the rate of 50 frames per second. The centres of the microrods were tracked from the recorded videos with a resolution of $\pm$20 nm~\cite{Zuhail2}. The autocorrelation and cross-correlation functions of the position fluctuations were computed using the Mathematica program.

\begin{figure}[!ht]
\center\includegraphics[scale=0.24]{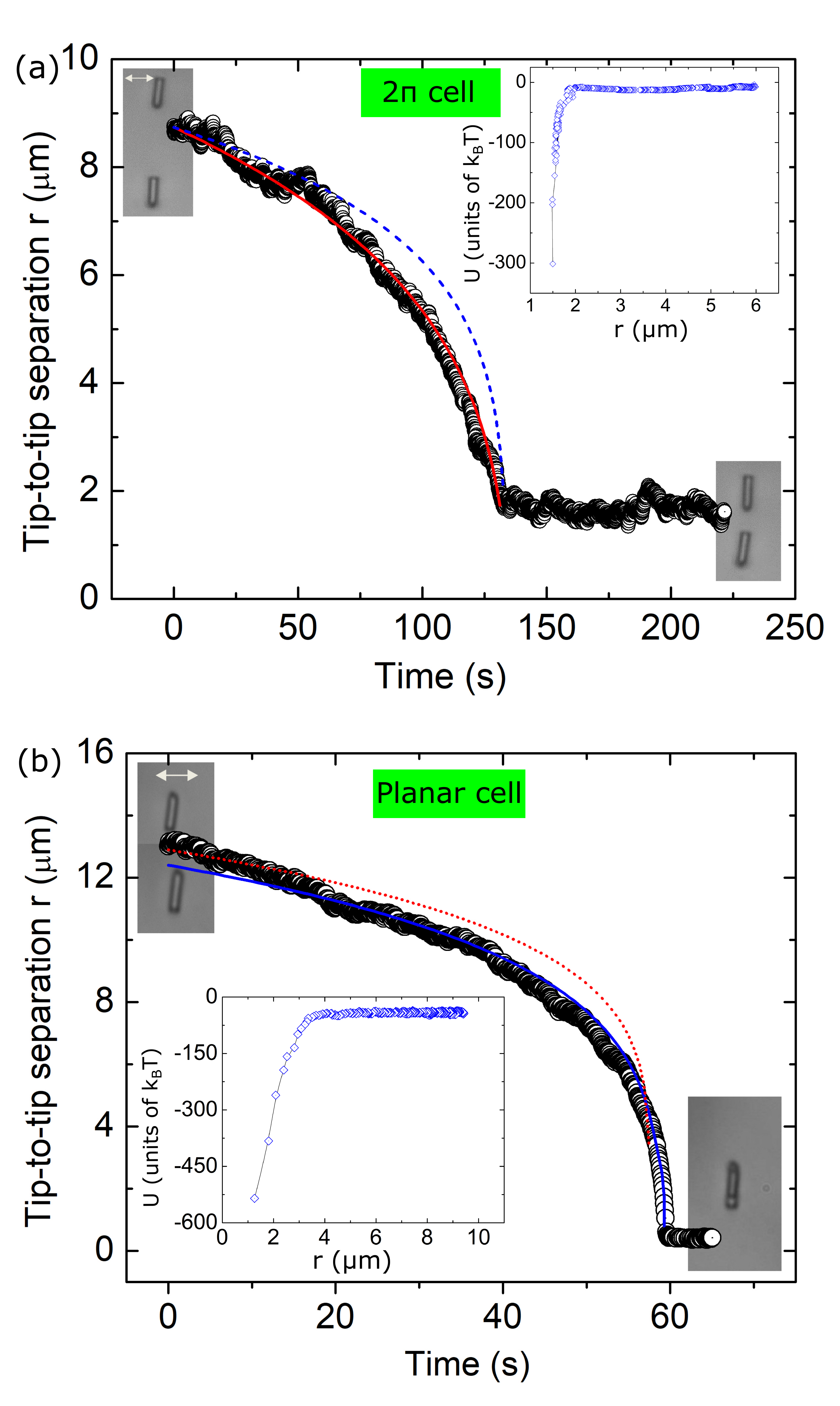}
\caption{(Color Online) (a) Time dependence of tip-to-tip separation of two interacting microrods in a $2\pi$-twisted cell (Movie S1). Solid red curve display the least square fit to the equation: $R(t)=(R_0^5-5\alpha_d t)^{1/5}$ for dipolar type interaction. (b)  Time dependence of tip-to-tip separation in a planar 5CB cell (without chiral dopant). Solid blue curve display the least square fit to the equation: $R(t)=(R_0^7-7\alpha_q t)^{1/7}$ for quadrupolar interaction. The insets show respective interaction potentials. The dashed blue curve in (a) and dotted red curve in (b) display fits to a quadrupolar and dipolar interactions, respectively.
 \label{fig:figure2} }
\end{figure}

\begin{figure}[!ht]
\center\includegraphics[scale=0.35]{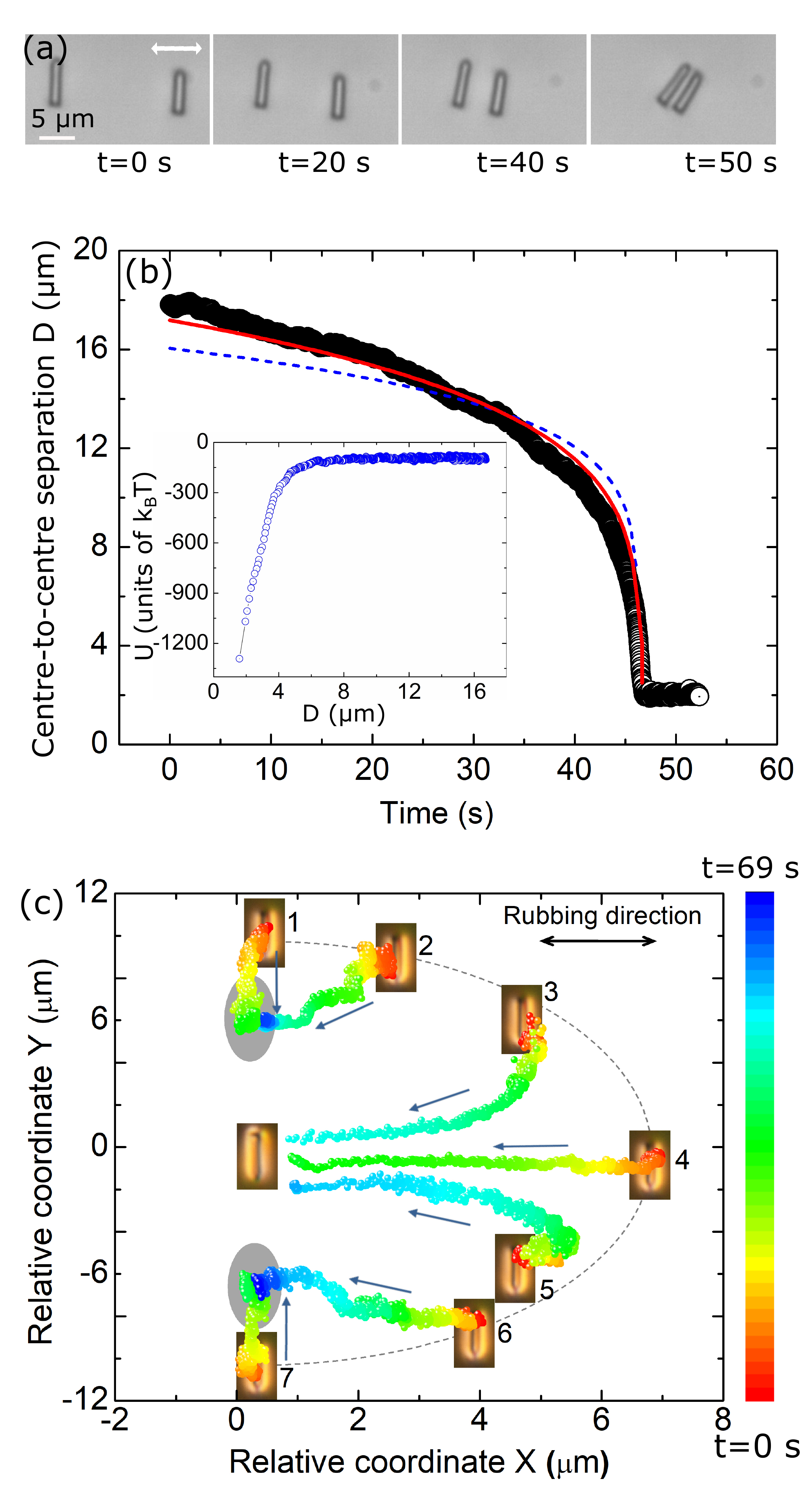}
\caption{(Color Online) (a) CCD images with elapsed time showing interacting microrods in a side-to-side configuration. (b) Time dependence  of the microrods separation in side-to-side configuration. Solid red curve display the best fit to equation: $R(t)=(R_0^5-5\alpha_d t)^{1/5}$ for dipolar interaction. Inset shows the interaction potential. The dashed blue curve display fits to a quadrupolar interaction.
(c) Time coded trajectories (relative coordinates) of microrods approaching from different starting positions in a 2$\pi$-twisted cell. By ``relative coordinate" we mean relative to the starting point of each trajectory. Grey shaded elliptical regions indicate where the approaching particle is bound with the central particle. 
\label{fig:figure3}}
\end{figure}

\section{Results and discussion}
\begin{figure}
\center\includegraphics[scale=0.3]{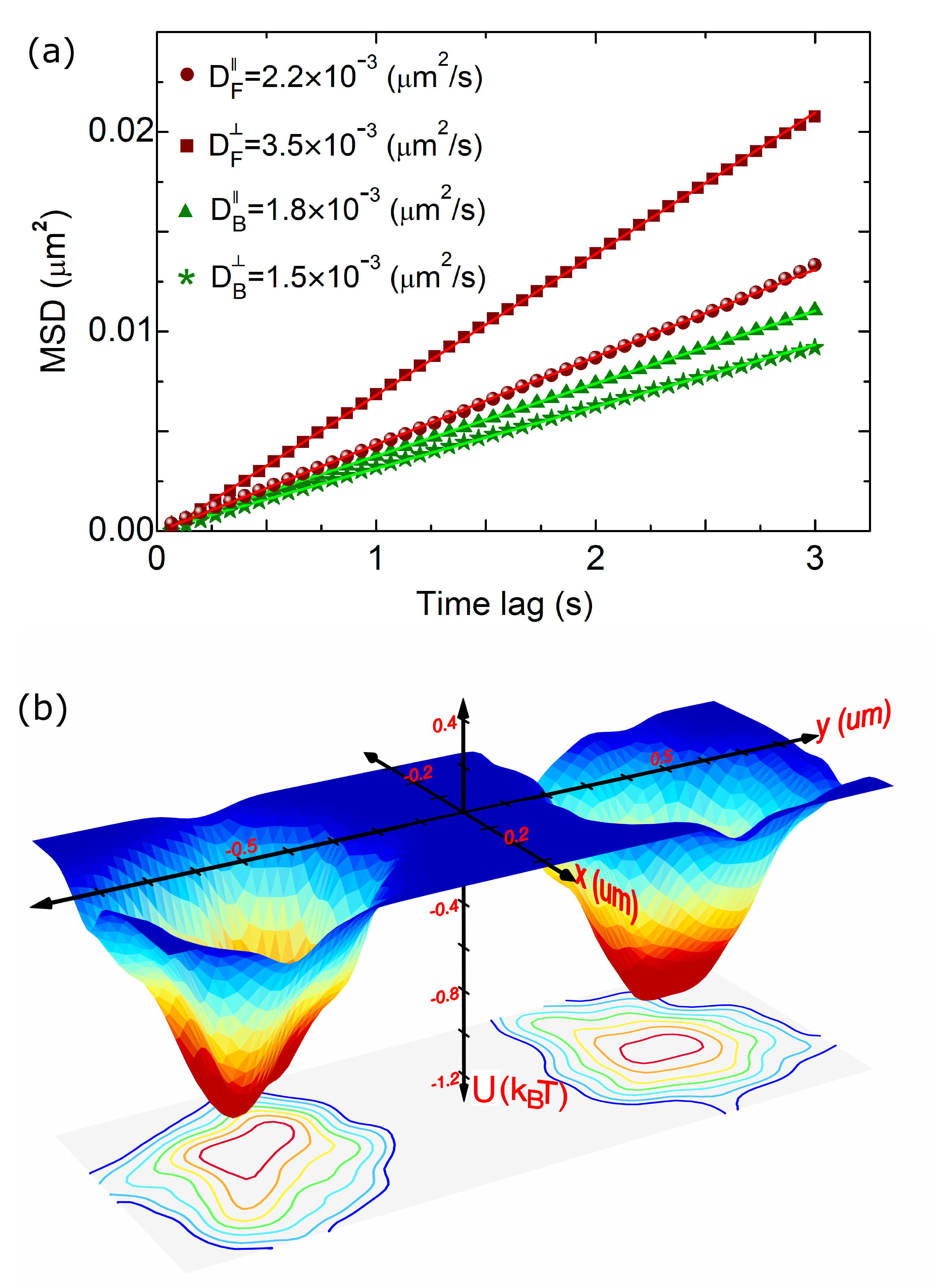}
\caption{(Color Online) (a) Mean squared displacements (MSDs) of a free microrod (red squares and circles) and a bound microrod (green triangles and stars) in a 2$\pi$-twisted cell. The subscripts F and B stands for free and bound microrods, respectively and the superscripts $||$ and $\perp$ indicate motions parallel and perpendicular to the rubbing direction. (b) Confining potentials of the bound microrods. Separation along the $y$-direction is normalized. 
\label{fig:figure4}}
\end{figure}

We work in the dilute concentration regime where most of the microrods are dispersed singly throughout the cell. To begin with, we study the LC-microrod dispersions in a 2$\pi$-twisted cell and observe that the majority of the microrods are orientated perpendicular to the rubbing direction (Fig.\ref{fig:figure1}(c)). Although the surrounding elastic distortion of the microrod is clearly visible (Fig.\ref{fig:figure1}(d)), the structure of the defect is unclear unlike the case of a microsphere dispersed in a planar, uniform nematic LC~\cite{Jampani}, wherein a single defect loop winds around the particle when the diameter ($D$) of the particle is comparable to the cell thickness and the twisting strength, $D/p$ is integral multiple of 1/2 where $p$ is the pitch. For microrods in $2\pi$-twisted cell, $D/p\simeq0.16$, which is much less than 1/2 and they have dipolar configuration (see later). 

Then we study the pair interaction of  microrods oriented perpendicular to the rubbing direction in a 2$\pi$-twisted cell. Two coaxial microrods were positioned at a certain distance with the help of the laser tweezers and allowed them to interact freely after switching off the laser. 
 Figure \ref{fig:figure1}(d) shows a few snapshots at different times while they are approaching to each other. Figure \ref{fig:figure2}(a) shows that the tip-to-tip separation decreases  with increasing time and eventually they are self-trapped at a certain separation forming a bound state. The mean separation between the microrods in the bound state (beyond 130 s) is 1.65 $\upmu$m and it remains fixed forever (see Fig.\ref{fig:figure8}(a) in Appendix A). It appears as if the microrods are trapped in the global energy minimum. Similar behaviour is also observed in cells with different chiralities and the tip-to-tip separation decreases moderately with increasing chirality as shown in the inset of Fig.\ref{fig:figure8}(b) (Appendix A). We also studied the interaction of two microrods in a planar 5CB without any chiral dopant. Figure \ref{fig:figure2}(b) shows the tip-to-tip separation of two microrods oriented perpendicular to the rubbing direction in a planar 5CB cell. In this case, the separation also decreases and finally one micrcorod goes under the other hence, the separation appears to be zero. Thus the equilibrium configuration of  the microrods in chiral NLCs is distinctly different than those in a planar nonchiral NLCs. 

The microrods in $2\pi$-twisted cell show a dipolar interaction which is evidenced from the fitting of the time dependence of the particle separation (Fig.\ref{fig:figure2}(a)) and given by $R(t)=(R_0^5-5\alpha_d t)^{1/5}$, where $\alpha_d$ is a constant and $R_0$ is the separation at $t=0$ s~\cite{Rasi}. On the other hand in planar 5CB cell (without chirality) they show quadrupolar type  interaction which is confirmed from the fitting of the time dependent separation to $R(t)=(R_0^7-7\alpha_q t)^{1/7}$ as shown in Fig.\ref{fig:figure2}(b). It is observed that in both cases $R(t)$ fits the data well with the adjustable fit parameters $\alpha_d= 1.0\times 10^3$$\upmu$m\textsuperscript{5}/s and $\alpha_q=3.2\times 10^{6}$$\upmu$m\textsuperscript{7}/s for dipolar and quadrupolar interactions, respectively. We are unable to fit the trajectory, shown in Fig.\ref{fig:figure2}(a) to a quadrupolar and the trajectory, shown in Fig.\ref{fig:figure2}(b) to a dipolar interactions, which are displayed by dashed curves. The interaction potentials shown in the insets  are calculated from $R(t)$. Since the system is highly overdamped (Reynolds number $<<1$) the Stokes drag force $F_S=6\pi L_{eff}\eta \partial R_{t}/\partial t=-F_{el}$, where  $L_{eff}$ is the effective length and $F_{el}$ is the elastic force~\cite{rod1}. The interaction potential $U$ is obtained by numerically integrating  $F_{el}$  over the distance moved $U=\int F_{el}.dR$. For this purpose we have measured independently the drag coefficients $\zeta_x$ and $\zeta_y$, parallel and perpendicular to the rubbing direction from the measurement of diffusion coefficients (see Fig.\ref{fig:figure4} (a)) and used an average value  $\zeta_a=2.2\times10^{-6}$ Kg/s for potential calculations.  
The minimum of the dipolar potential energy in $2\pi$-twisted cell (inset of Fig.\ref{fig:figure2}(a)) is 350 k\textsubscript{B}T which is slightly lesser than that of the quadrupolar interaction energy (540 k\textsubscript{B}T ) in the planar 5CB cell (Fig.\ref{fig:figure2}(b)).

We also studied the interaction of microrods in a side-to-side configuration as shown in Fig.\ref{fig:figure3}(a). In this case the dipolar microrods have an antiparallel orientation. They are attracted and finally settled at a small surface-to-surface separation (600 nm).  The time dependence of the centre-to-centre separation is shown in Fig.\ref{fig:figure3}(b). It is well fitted to $R(t)=(R_0^5-5\alpha_d t)^{1/5}$ for a dipolar type interaction with fit parameter $\alpha_d= 6.4\times 10^3$$\upmu$m\textsuperscript{5}/s. Here, we are unable to fit the trajectory to a quadrupolar interaction as shown by dashed curve. The interaction potential for this configuration is shown in the inset. The minimum of the potential energy is about 1300 K\textsubscript{B}T which is much larger than that of the tip-to-tip configuration (inset to Fig.\ref{fig:figure2}(a)).
The overall anisotropy of the pair interaction was studied by looking at the relative trajectories of two microrods released from different starting positions as shown in Fig.\ref{fig:figure3}(c). Microrods starting from positions 1,2,6 and 7 are assembled to form a bound state with a larger separation than  the microrods released from the starting positions 3,4,5.
 \begin{figure}[!ht]
\center\includegraphics[scale=0.35]{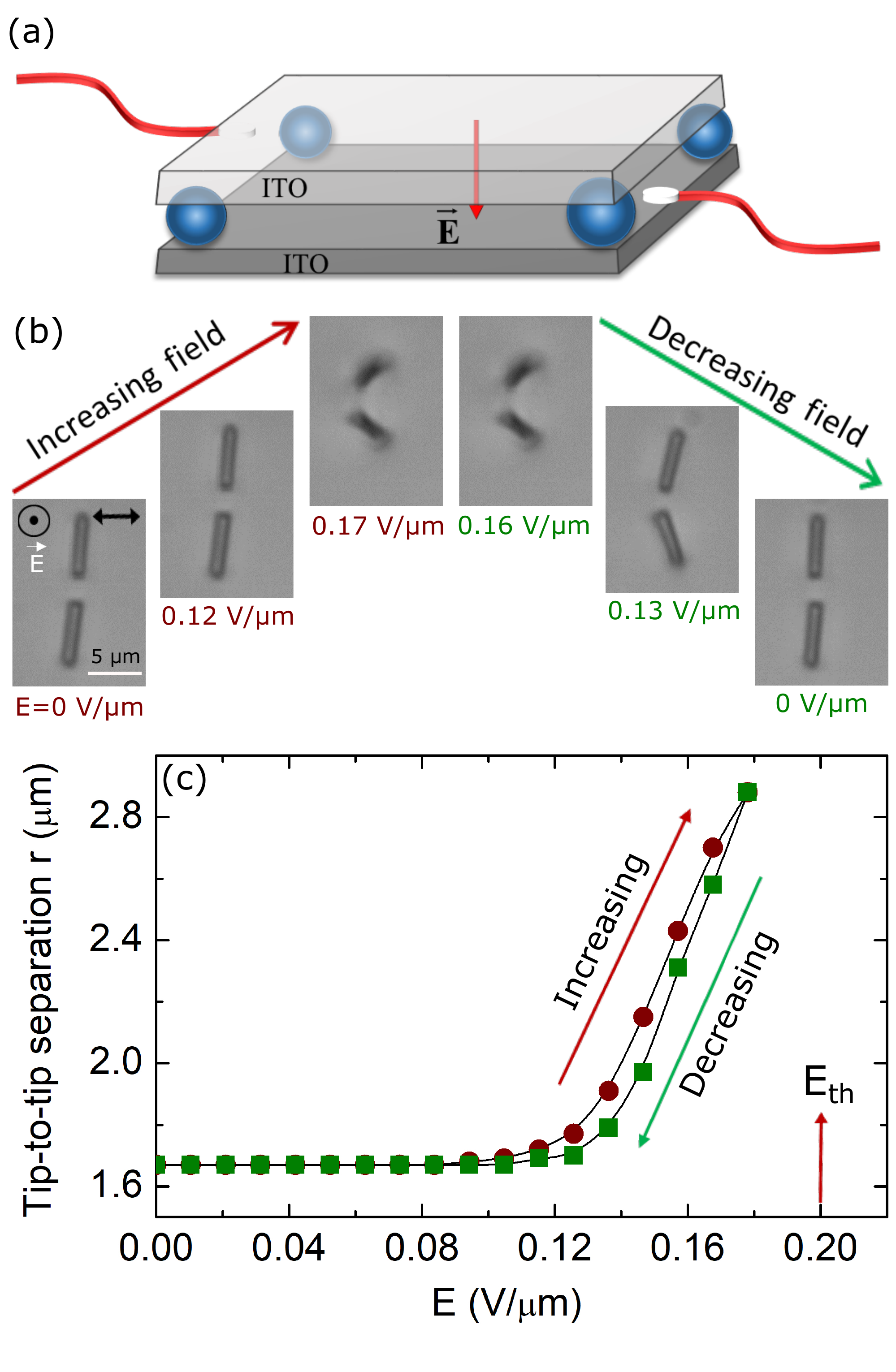}
\caption{(Color Online) (a) Diagram of the cell for the electric field experiments. The direction of the electric field between the ITO plates is perpendicular to the director. (b) CCD images with elapsed time showing the effect of ac electric field ($f=1$ kHz) on the bound state in  a 2$\pi$-twisted cell (Movie S3)~\cite{sup} (c) Variation of tip-to-tip separation with increasing (circles) and decreasing (squares) electric field. The vertical arrow indicates the Freedericksz threshold field (E\textsubscript{th}).
\label{fig:figure5}}
\end{figure}
 
 It has been reported that the spherical microparticles in chiral NLCs exhibit metastable states due to the screening of chiral interaction in which two microspheres are temporarily held at a certain separation for a short duration and eventually they are attracted and joined together~\cite{Jampani}. We have also performed experiments on spherical microparticles and demonstrated metastable state in $\pi$ and $2\pi$-twisted cells (Fig.\ref{fig:figure9}, Appendix B). The striking result of our experiment is that the microrods are self-trapped at a certain separation  forming a stable bound state. The origin of the bound state can be understood from the competing effect of the electrostatic and elastic interactions. The surface of the silica microrods are negatively charged~\cite{rod}. 
To show the effect of surface charge we heated the sample to the isotropic phase where the elastic interaction between them is absent and pushed the microrods towards each other with the help of laser tweezers (see Movie S2)~\cite{sup}. The microrods stay away from each other against the impelling force, demonstrating an electrostatic repulsion between them due to the surface charge. We measured the total charge of the individual microrods from the electrophoretic mobility of the particles in a chiral NLC under the dc electric field. The details of the measurement are presented in Fig.\ref{fig:figure10} (Appendix C). The total charge of a microrod is $Q\simeq -(1034\pm24) e$ ($e=1.6\times10^{-19}$ C) and the corresponding Coulomb energy between the two charged microrods at a centre-to-centre separation of 8 $\upmu$m is $2.9\times 10^3$ k\textsubscript{B}T. Thus, electrostatic potential energy is about one order of magnitude larger than the elastic energy (300 k\textsubscript{B}T) of the microrods (see inset of Fig.\ref{fig:figure2}a)). Hence, the metastable state becomes a bound state due to the electrostatic repulsion.
 

\begin{figure}[!ht]
\center\includegraphics[scale=0.4]{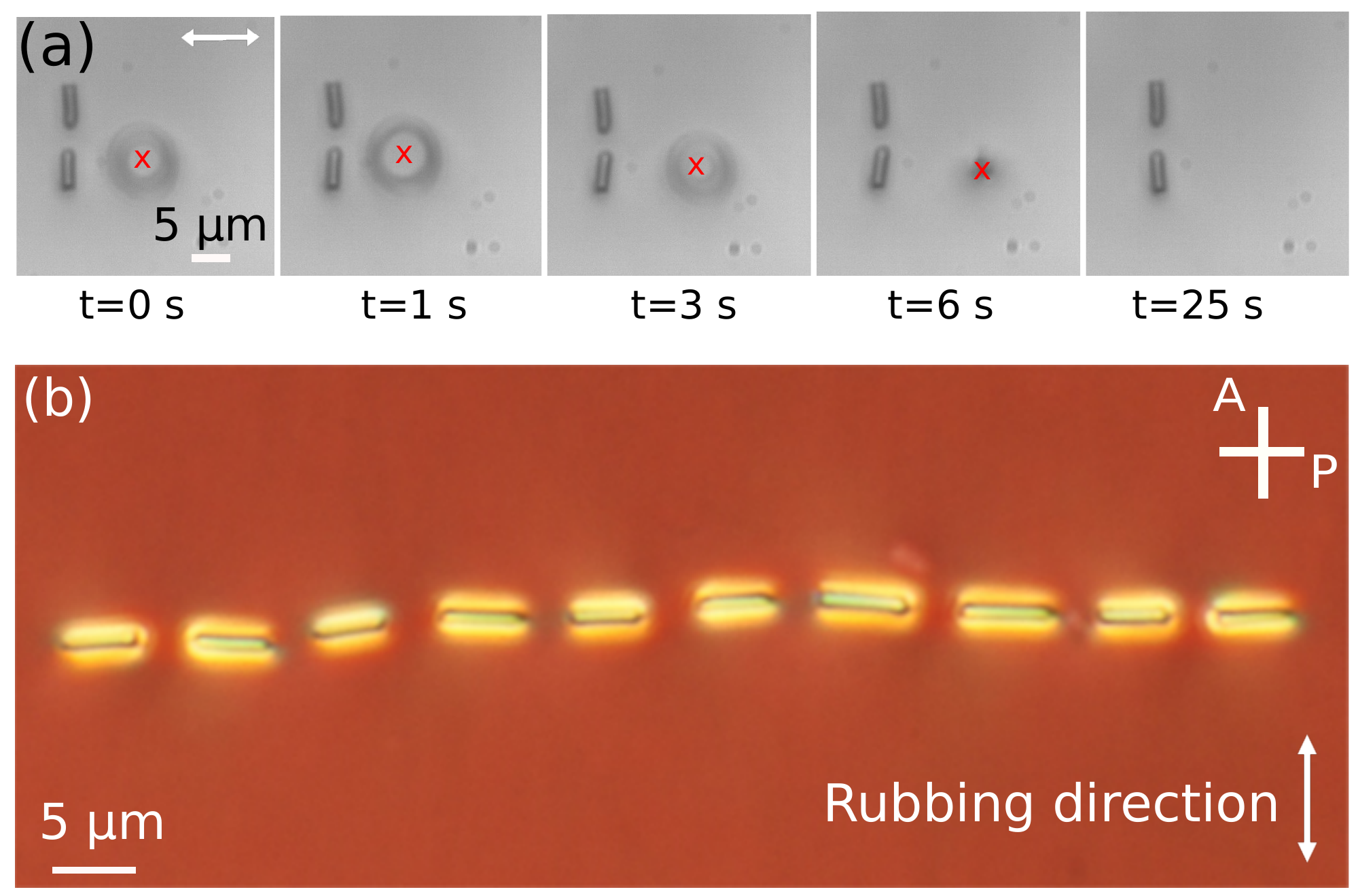}
\caption{(Color Online) (a) CCD images with elapsed time showing the effect of elastic distortion created by the laser tweezers on the bound state in a 2$\pi$-twisted cell. (Movie S4)~\cite{sup} (b) A linear chain of 10 microrods forming successive bound states. They were guided to assemble with the help of the laser tweezers setup. 
\label{fig:figure6}}
\end{figure}

 In the bound state the Brownian motions due to thermal agitation of the microrods are clearly observed (see Movie S1)~\cite{sup}. The self-diffusion coefficients $D_{B}^{||,\perp}$ of a bound microrod and $D_{F}^{||,\perp}$ of a free or isolated  microrod (where $||$ and $\perp$ refers in relation to the rubbing direction) were obtained from the mean squared displacement. Figure \ref{fig:figure4}(a) shows that $D_{B}^{\perp}\simeq 0.4 D_{F}^{\perp}$ and $D_{B}^{||}\simeq 0.8 D_{F}^{||}$ i.e., the diffusion coefficients in the bound state are reduced considerably than that of a free microrod as expected. In the isotropic solvent the proximity of the surface due to the sedimentation often alters the Brownian motion and hence the friction coefficients~\cite{lib}. However, in NLCs, sedimentation of microparticles is prevented by thermal fluctuations and  the elasticity of the medium as the gradients of the director field push the particles away from the bottom~\cite{olegprl}. Hence, the Brownian motion and the estimated friction coefficients of the microrods are not influenced by the substrates. It may be mentioned that the anisotropic particles in an isotropic solvent like, water have both positional and orientational degrees of freedom~\cite{sci}. However, in our system the orientational fluctuations of the microrods are reduced compared to that in water due to the nematic order.

We have measured the effective confining potentials from the study of the Brownian statistics of the microrods in the region accessible by thermal agitations. In equilibrium the probability density of the particle position is given by $p(x)=C\exp[-U(x)/k_{B}T]$, where $C$ is the normalisation constant and $U(x)$ is the confining potential~\cite{Osterman}.  The potential is obtained from the normalised histogram of the confined particle's positions and given by $U(x)=-ln[p(x)]$. Figure \ref{fig:figure4}(b) shows the confining potentials of the microrods at a normalised centre-to-centre distance. The depth of the potential energy is $\sim k_{B}T$ and the shape of the potentials is asymmetric as expected (see contour plots). Assuming the potentials are harmonic, the effective stiffness constants $k_x$ (along rubbing direction) and $k_y$ (perpendicular to the rubbing direction) are given by $k_{x}=4.3\pm 0.1\times10^{-6}$ N/m and $k_{y}=4.8 \pm 0.1\times10^{-6}$ N/m. The stiffness constants of the confining potentials are comparable to that of the low power optical traps~\cite{Jens,trap}.   

To test the robustness of the bound state we applied an external ac electric field that perturbs the position and orientation of the microrods (Movie S3). The field is applied between the two ITO (indium-tin-oxide) electrodes and its direction is perpendicular to the director as shown in Fig.\ref{fig:figure5}(a).  Figure.\ref{fig:figure5}(b) shows some snapshots at different field strengths. The maximum field applied  (0.18 V/$\upmu$m) is well below the Freederickzs threshold value (0.2 V/$\upmu$m) so that there is no director reorientation. As shown in Fig.\ref{fig:figure5}(c) the microrods are displaced above the field 0.12 V$/\upmu$m,  and they also tilt simultaneously in the plane as well as out of the plane.  The initial state is recovered when the electric field is reduced to zero. The tilting of the microrods along the field direction beyond a particular field is due to the competing effect of electric field-induced torque and hydrodynamic torque~\cite{nano,ras1}. We further checked the stability of the bound state by applying a mild mechanical force with the help of the laser tweezers as shown in Fig.\ref{fig:figure6}(a). It is observed that the bound state is also stable against the elastic distortion created by the trap surrounding the microrods (Movie S4). With the help of the laser tweezers, we have created an assembly of several microrods in the form of a linear chain as shown in Fig.\ref{fig:figure6}(b). Here, the microrods are  connected through successive bound states. The chain is a little zigzag which could be due to a slight variation in the length and diameter of the microrods. Nevertheless, the chain is highly stable against thermal fluctuations and mild perturbing force applied by the laser tweezers (Movie S5) and remains unchanged even after a few days. It may be mentioned that although one end of the microrods is round-shaped the bound state formation does not depend on the shape of the facing ends. 

\begin{figure}[!h]
\center\includegraphics[scale=0.22]{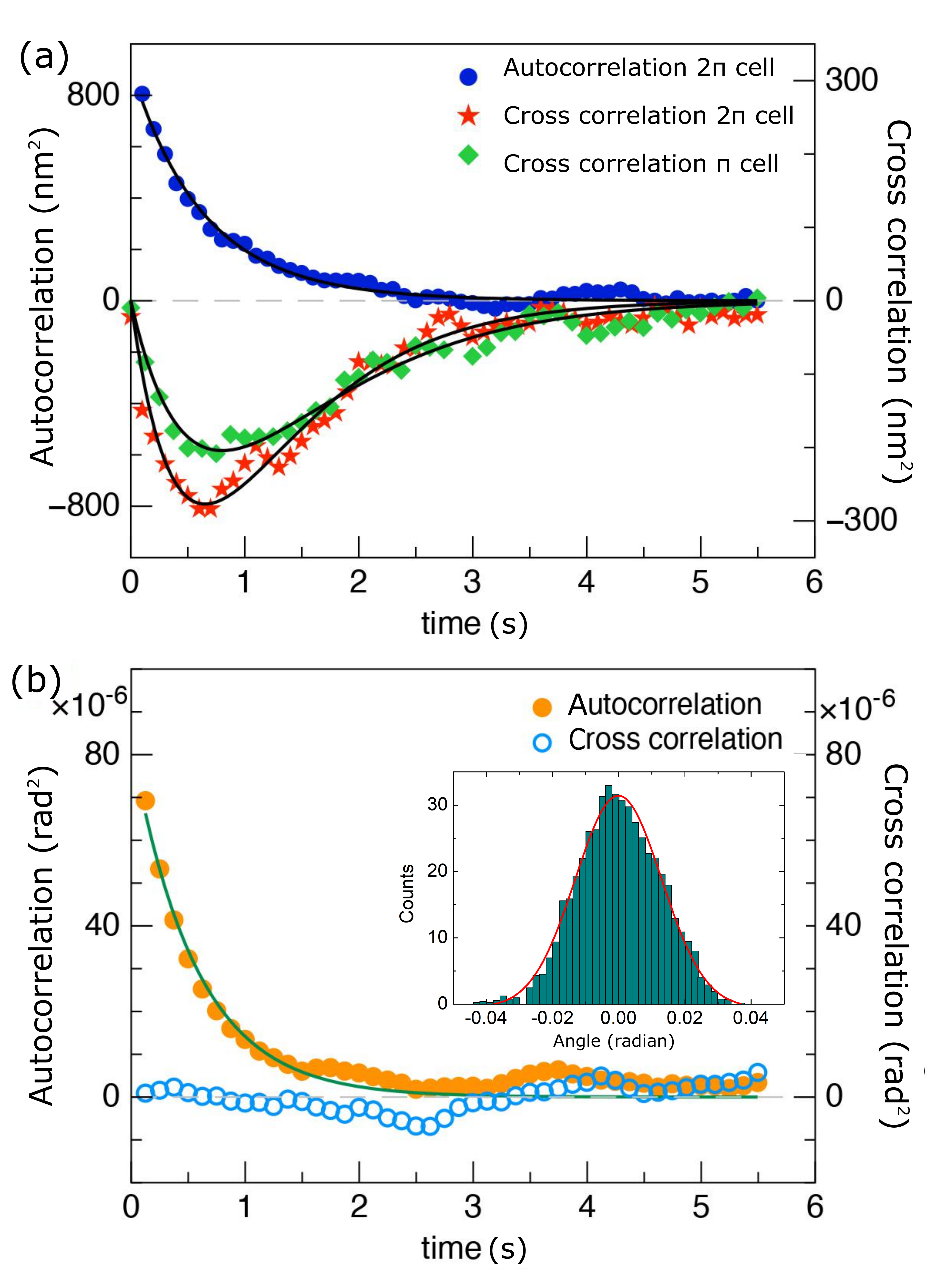}
\caption{(Color Online) (a) Longitudinal autocorrelation $\langle R_{1}(t)R_{1}(0)\rangle$  (blue circles) and cross-correlation $\langle R_{1}(t)R_{2}(0)\rangle$ (green diamonds and red stars) functions of the  microrods in $\pi$ and 2$\pi$-twisted cells.  Solid lines represent the least square fits to Eq.(1) and Eq.(2) for auto and cross correlation functions. (b) Autocorrelation function $\langle\theta_{1}(t)\theta_{1}(0)\rangle$ (solid circles) and cross-correlation function $\langle\theta_{1}(t)\theta_{2}(0)\rangle$ of orientational fluctuations  (open circles) of two microrods in 2$\pi$-twisted cell. Solid line indicates the least square fit to $\langle\theta_{1}(t)\theta_{1}(0)\rangle$. Only every third of the experimental data points is shown. The inset shows the histogram of orientational deviation.
\label{fig:figure7}}
\end{figure}

In what follows we study the correlated thermal fluctuations of the microrods in the bound state. We measure the position of the particles and calculate the autocorrelation and cross-correlation functions. Figure \ref{fig:figure7}(a) shows that the autocorrelation function decays exponentially as expected whereas a strong time delayed-anticorrelation is observed with a pronounced minimum. Similar anti-correlation of a pair of optically trapped microspheres in aqueous solution have been reported and it was explained on account of their hydrodynamic interactions~\cite{Jens,Bar}. In oder to explore possible hydrodynamic coupling between the two microrods we have analysed our results following the method discussed in Refs~\cite{Jens,Bar}.
  In the framework of Langevin dynamics for a low Reynolds number system the autocorrelation and the cross-correlation functions of position vectors of the two particles along the y-direction (perpendicular to the rubbing direction) is expressed as~\cite{Jens}

\begin{align}
\langle R_{1}(t)R_{1}(0)\rangle 
	&=\langle R_{2}(t)R_{2}(0)\rangle\nonumber\\
	&= \frac{k_BT}{2k_y}(e^{-t(1+\epsilon)/\tau}+e^{-t(1-\epsilon)/\tau})
	\label{eqn:acf}
\end{align}
and 
\begin{align}
\langle R_{1}(t)R_{2}(0)\rangle 
	&=\langle R_{2}(t)R_{1}(0)\rangle\nonumber\\
	&= \frac{k_BT}{2k_y}(e^{-t(1+\epsilon)/\tau}-e^{-t(1-\epsilon)/\tau})
	\label{eqn:ccf}
\end{align}
where $\tau$ is the relaxation time and $\epsilon$ is a dimensionless parameter. The autocorrelation and cross-correlation functions measured in $\pi$ and $2\pi$-twisted cells and the theoretical fits to Eq.(1) and Eq.(2) are shown in Fig.\ref{fig:figure7}(a). The parameters obtained from the fittings are given by $\tau=0.64\pm 0.01$ s and $\epsilon=0.10\pm0.01$.  It means the hydrodynamic interaction is  asymmetric as one  microrod moves it tends to drag the other and the correlated fluctuations relax faster than the anticorrelated fluctuations. We directly estimate $\epsilon$ from the particle separation and $\tau$ from the effective confining potentials. 
  $\tau$ is given by the ratio of the friction coefficient $\zeta_y$ and the stiffness constant $k_y$ of the potential i.e., $\tau=\zeta_y/k_y$. The parameter $\epsilon$ describes the ratio of the mobility of the particle and the strength of the hydrodynamic coupling and expressed as $\epsilon \approx 3a/2r$, where $r$ is the mean separation and $a$ is the radius of the particle~\cite{Jens}. Using Stokes-Einstein relation and measured diffusion coefficient (Fig.\ref{fig:figure4}(a)), we obtain $\zeta_y=2.8\times10^{-6}$ Kg/s. Taking  $k_y=4.8\times10^{-6}$ N/m from the confining potentials (Fig.\ref{fig:figure4}(b)), the radius of the microrod $a=0.4 ~\upmu$m and the centre-to-centre separation $r\simeq7~\upmu$m, the calculated parameters are $\tau=0.58$ s and $\epsilon\simeq 0.09$. These parameters are in good agreement with those obtained from the fitting of the autocorrelation and cross-correlation functions. It is also observed from Fig.\ref{fig:figure7}(a) that the depth of the minimum in $2\pi$-twisted cell is higher than that of $\pi$-twisted cell. In $2\pi$-twisted cell the particles are closer than $\pi$ cell (inset of Fig.\ref{fig:figure8}(b)) hence, the result is consistent with the  prediction that the strength of the hydrodynamic coupling is inversely proportional to the separation~\cite{Jens}. We would like to mention that the above model for spherical particles dispersed in an isotropic solvent is adapted for NLCs, assuming that the nematic director remains uniform over the length scale of inter particle separation hence, the hydrodynamic contribution to the director mode is negligible. This could be reasonable as the NLC is aligned by confining surfaces and the director is fairly uniform in the gap between the two tips of the two microrods [Fig.\ref{fig:figure1}(d)]. Hydrodynamic interaction has been considered in two-point particle tracking microrheology in NLCs~\cite{rheo}. Theoretically it could be  interesting to investigate the contribution of the director mode on the two particle auto and cross correlation functions.
  
In the bound state, the orientational fluctuations of the microrods are clearly observed. The orientational fluctuations $\theta_1(t)$ and $\theta_2(t)$ of the long axes of the microrods are measured with respect to their mean orientations. The histogram of the orientational fluctuations of a microrod in the bound state is a Gaussian as shown in the inset of Fig.\ref{fig:figure7}(b). We measured the autocorrelation $\langle\theta_{1}(t)\theta_{1}(0)\rangle$ and cross-correlation $\langle\theta_{1}(t)\theta_{2}(0)\rangle$ functions of the orientational fluctuations.  Figure \ref{fig:figure7}(b) shows that the autocorrelation function decays exponentially and it can be well fitted to $\langle\theta_{1}(t)\theta_{1}(0)\rangle$$\sim e^{-t/\tau_o}$ with a decay constant $\tau_{o}=56$ ms. This value is comparable to the effective director relaxation time of 5CB liquid crystal ($\simeq 50$ ms) measured by the dynamic Freedericksz transition~\cite{note}. The cross-correlation function is almost zero, suggesting that the orientational fluctuations are not coupled hydrodynamically. \\
We make some necessary comments here. The obtained correlation relaxation time of hydrodynamic interaction in our chiral NLCs is $\tau=0.64$ s, which is 15 times larger than that of a microsphere in water (43 ms)~\cite{Jens}. This is expected due to the larger viscosity of the liquid crystal ($\eta^{5CB}_{||}\simeq 20$ mPas)~\cite{chim} with respect to that of the water ($\eta_W\simeq 1$ mPas). Although the theoretical technique adapted here is described for hydrodynamic interactions in low Reynolds number (R\textsubscript{e}) fluid systems, we expect it to be valid as R\textsubscript{e} of NLCs is much smaller than 1 (R\textsubscript{e}$\sim10^{-4}-10^{-5}$)~\cite{oleg,ours}. Here, we have demonstrated the bound state of charged microrods in chiral NLCs. However, similar experiments could be performed on particles with any shape and one can expect the formation of the bound state, provided the elastic and electrostatic forces are appropriately balanced in some direction.  \\

\section{Conclusion} 
 In conclusion, two collinear charged silica microrods aligned perpendicular to the rubbing direction in chiral nematic liquid crystals are self-trapped due to the competing effect of elastic attraction and Coulomb repulsion forming a colloidal bound state. The bound state is highly stable and robust against the influence of external electrical and mechanical perturbations. In the bound state the positional fluctuations of the microrods are hydrodynamically coupled whereas their orientational fluctuations are uncoupled. The correlation relaxation time of hydrodynamic interaction in chiral nematic liquid crystals is more than one order of magnitude larger than that in aqueous suspensions. 
 Apart from the interplay of elasticity and defects, the hydrodynamic interaction in chiral NLCs play an important role in colloidal self-assembly. The abundance of new shape anisotropic particles and their unusual topological properties in chiral NLCs widen the scope of further studies.\\

{\bf Acknowledgments:}
SD  acknowledges the support from the Department of Science and Technology, Govt. of India (DST/SJF/PSA-02/2014-2015),  DST-PURSE-II and UoH (UoH/IoE/RC1-20-010). Ravi Kumar Pujala acknowledges DST for INSPIRE Faculty Award Grant [DST/INSPIRE/04/2016/002370] and Core Research Grant [CRG/2020/006281] from SERB. P.S. acknowledges IISER Tirupati for the research funding. M.R.M acknowledges CSIR fellowship.  We acknowledge useful discussion with Sriram Ramaswamy. 

\subsection*{\bf{APPENDIX A: \uppercase{Bound state of microrods in twisted cells}}}
\begin{figure}[!ht]
\center \includegraphics[scale=0.32]{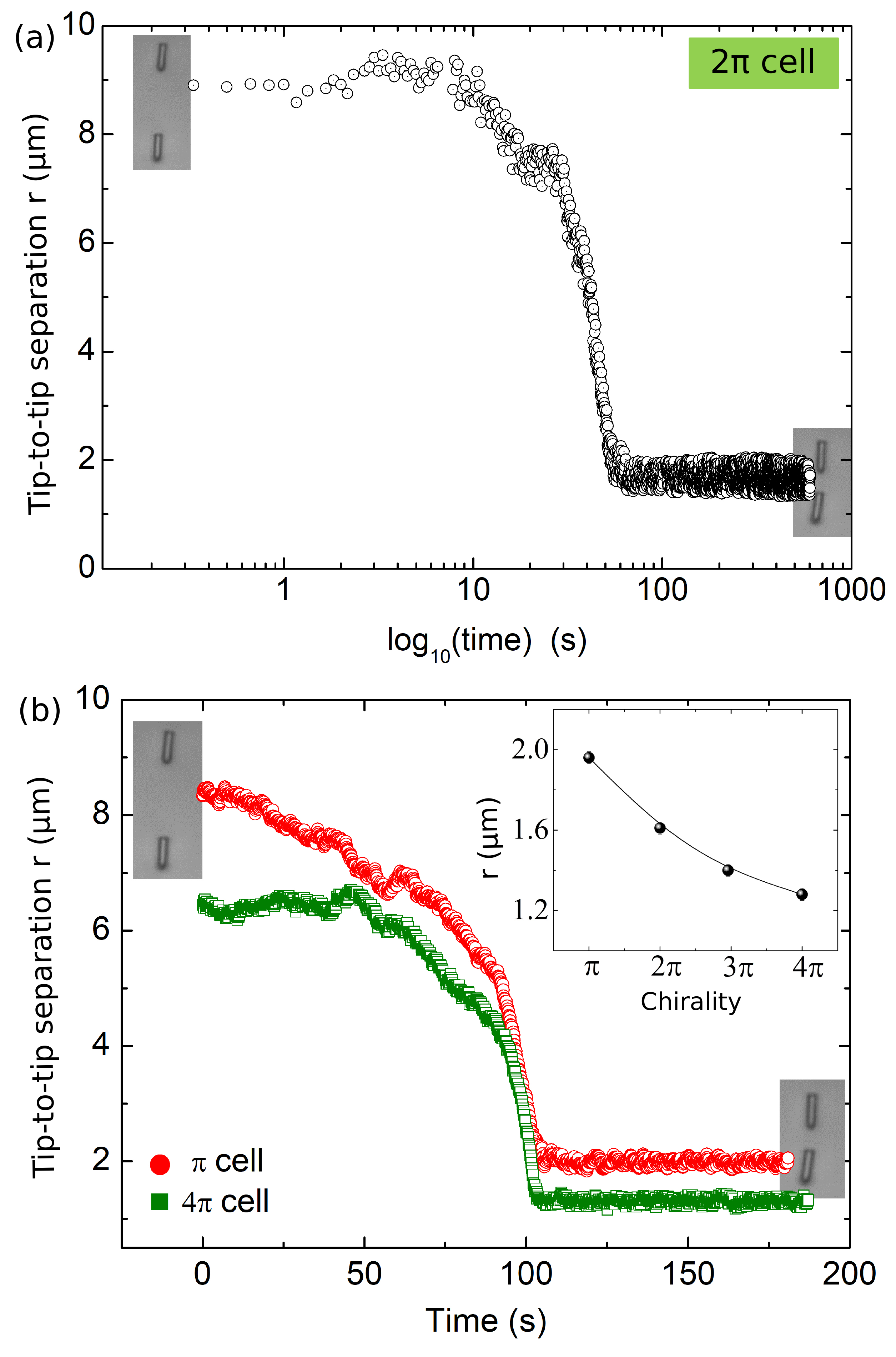}
\caption{(Color Online) (a) Tip-to-tip separation between a pair of microrods in 2$\pi$-twisted cell with time (in semi-log scale). The data is presented here for about 10 minutes but the observation was made for 3 to 4 hrs to confirm that the bound state is stable. (b) Tip-to-tip separation in $\pi$ (red circles) and $4\pi$-twisted (green squares) cells. Inset shows the tip-to-tip separation with chirality.}
\label{fig:figure8}
\end{figure}

To show the stability of the bound state against the thermal fluctuations we observed the microrod pair for a longer duration. Figure \ref{fig:figure8}(a) shows the tip-to-tip separation of the microrods in a 2$\pi$-twisted cell for about 10 minutes. Further the colloidal pair was observed for several hours (3 to 4 hrs) to ensure that the bound state is stable. Figure \ref{fig:figure8}(b) shows the end-to-end separation in two different cells namely, $\pi$ and $4\pi$-twisted cells. It is observed that the equilibrium separation in the bound state is lower in $4\pi$-twisted cell than that in the $\pi$-twisted cell. Inset shows that the equilibrium separation decreases with increasing chirality.

\subsection*{\bf{APPENDIX B: \uppercase{Metastable state of microspheres in a 2$\pi$-twisted cell }}}
\begin{figure}[!ht]
\center \includegraphics[scale=0.35]{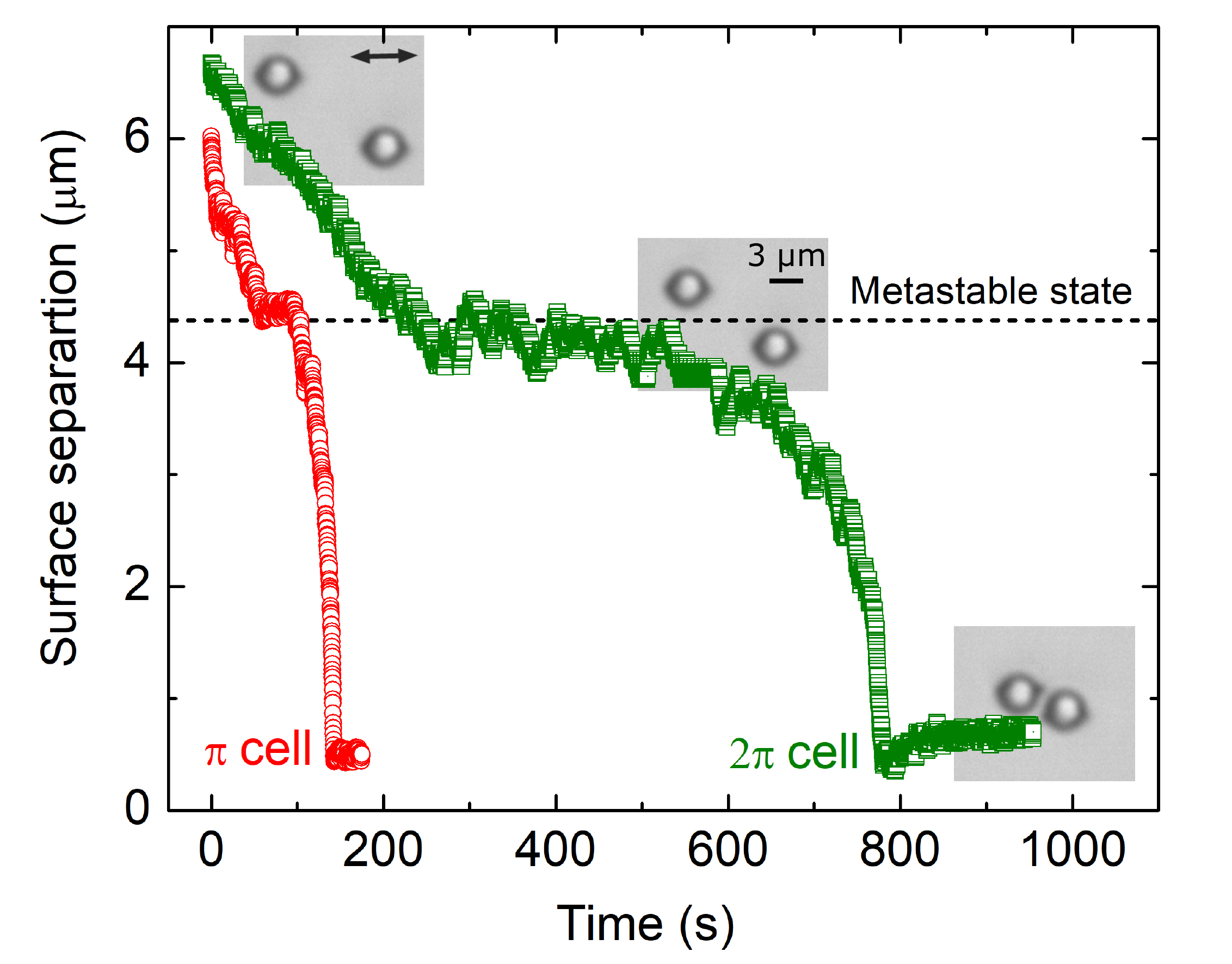}
\caption{(Color Online) Surface-to-surface separation of two spherical DMOAP coated silica microspheres of diameter 3 $\upmu$m in $\pi$ (red circles) and 2$\pi$-twisted  (green squares) cells. Ratio of the diameter to pitch, $D/p=1/2$ and 1 for $\pi$ and $2\pi$-twisted cells, respectively. Note that eventually the particles are attracted to join.}
\label{fig:figure9}
\end{figure}

To bring out the contrast of the responses of the microrods and microspheres we also performed similar experiments on spherical microparticles in two chiral cells following the report of Jampani \textit{et al.}~\cite{Jampani}. We used DMOAP coated silica microspheres of diameter $3~\upmu$m and studied them in $\pi$ and and $2\pi$-twisted cells. Figure \ref{fig:figure9} shows that there is a metastable state at which the surface-to-surface separation between the particles remains almost constant for some duration. For example, in $2\pi$-twisted cell the duration is about 260 s and it is much shorter for $\pi$-twisted cell (38 s).

\subsection*{\bf{APPENDIX C: \uppercase{Estimation of total charge $q$}}}
\begin{figure}[!ht]
\center \includegraphics[scale=0.24]{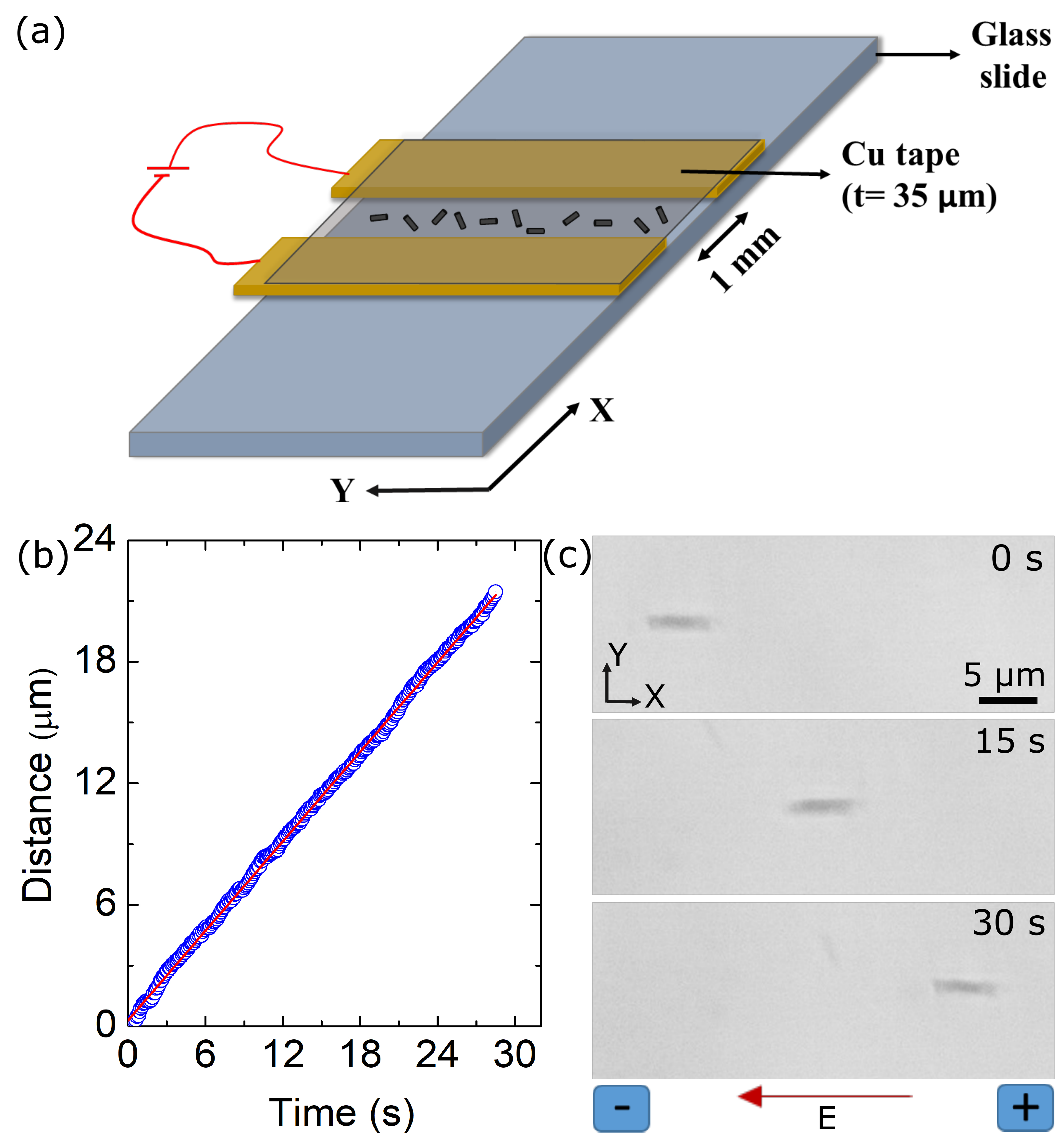}
\caption{(Color Online) Measurement of total charge $Q$ of DMOAP coated microrods in a chiral NLC. (a) Schematic diagram of the cell used in the experiment. (b) Variation of displacement of the particle position with time when a dc field of 10 V is applied between the copper (Cu) tapes, separated by 1.0 mm distance.  Solid red line is a linear fit to the data. (c) CCD images showing the motion of a microrod under the field. Note that the microrod moves from the negative to the positive electrode, indicating they are negatively charged.}
\label{fig:figure10}
\end{figure}

We have measured the total charge $Q$ of individual microrods from the electrophoretic motion under the action of a dc electric field \textbf{E}~\cite{ivan1}. For this experiment we prepared planar cells for applying in-plane electric field. A schematic diagram of the cell is shown in Fig.\ref{fig:figure10}(a). Two copper tapes (Cu) of thickness 35 $\upmu$m and width 5 mm are placed at a separation  $d=1$ mm on a glass plate. The top side is covered by a thin glass plate and the cell is sealed by a UV curable adhesive (NOA-81, Norland). The copper tapes also act as spacers. A very dilute dispersion of the microrods is injected into the cell. The electrodes are connected to an external dc power supply (TD3202M, Aplab). When a dc electric field of magnitude  $U$=10 V is applied, the microrods move from the negative to the positive electrode and the direction of motion is reverted by changing the polarity of the field. It indicates that the microrods are negatively charged. This was further confirmed by  measuring the zeta potential of the microrods $\Psi_0\simeq-13\pm1$ mV in ethanol using the light scattering technique (Litesizer 500, Anton Paar). The motion of the microrods is recorded by the CCD camera fitted with the inverted microscope and the position (centre) of the microrods is tracked. A few CCD images of the moving particle with elapsed time is shown in Fig.\ref{fig:figure10}(c). The velocity $v$ of the microrods under the dc field is obtained from the time dependent position as shown in Fig.\ref{fig:figure10}(b). Since the Reynold's number R\textsubscript{e}$\ll 1$, the total charge $Q$ can be obtained by balancing the viscous drag force $F_S=\zeta_a v$ with the electrostatic force $F_Q=QE$ acting on the microrods, where the average friction coefficient $\zeta_a=k_BT/D$, $D$ being the diffusion coefficient. Equating these two forces we get the total charge, $Q=\zeta_a vd/U$. The average friction coefficient is obtained from the average diffusion coefficients (Fig.\ref{fig:figure4}(a)) and given by $\zeta_a=2.2\times10^{-6}$ Kg/s. The velocity of the particles under dc field (10 V) is obtained from the displacement-time graph and is given by $v=0.76\pm0.1~\upmu$m. The total charge of a microrod is estimated to be $Q\simeq -(1034\pm24) e$, $e=1.6\times10^{-19}$ C is an elementary charge.

\end{document}